\documentstyle[12pt,a4]{article}

\begin{document}
\vspace{-1.5cm}

\begin{center}
{\Large \bf A condition for first order phase transitions in}

{\Large \bf quantum mechanical tunneling models}

\vspace{2cm}

{\bf  H. J. W. M\"uller--Kirsten{\footnote[1]{Email:mueller1@physik.uni-kl.de}},
D. K. Park{\footnote[2]{Email:dkpark@chep5.kaist.ac.kr, permanent address: Department
of Physics, Kyungnam University, Masan 631--701, Korea}} and J. M. S. Rana{\footnote[3]
{Email:jmsrana@physik.uni-kl.de}}}

{\it Department of Physics, University of Kaiserslautern,}

{\it 67653 Kaiserslautern, Germany}
\end{center}          

\vspace{1in}

\centerline{\large{\bf Abstract}}
A criterion is derived for the determination of parameter domains of first order phase transitions in
quantum mechanical tunneling models.  The criterion is tested by application to various
models, in particular to some which have been used recently to explore spin tunneling
in macroscopic particles. In each case agreement is found with previously
heuristically determined domains.

\newpage

\section{Introduction}
It is well known that at temperatures close to zero, decay rates of metastable states are
determined by quantum tunneling processes whose dynamics is described by classical
configurations such as vacuum or periodic bounces, but that with increasing temperature
thermal activation becomes more and more important, and beyond some critical
or crossover temperature $T_c$ becomes the decisive mechanism.  It has been realised recently
that this change or crossover can be regarded as a transition from quantum to classical
behaviour which in turn can be looked at like a phase transition in e.g. the thermodynamics
of gases.  This is a fascinating topic which has attracted considerable interest recently
and application particularly in the area of macroscopic spin systems.  The general
idea applies also to other cases where instead of a decay rate a shift in energy is involved, such as a
level splitting in quantum mechanics which is described by vacuum or periodic
instantons.  

The characteristic way in which phase transitions appear in quantum mechanical tunneling
processes has been worked out in ref.\cite{1}.  In particular, a sharp
first order transition is there shown to appear in the plot of action $S$ versus 
temperature $T$ which is very analogous to that of the free enthalpy versus pressure of a 
van der Waals gas, whose equation of state plotted as pressure versus volume corresponds to the
plot of the period $P(E)$ (of the periodic bounce or instanton) versus energy in the
consideration of quantum mechanical systems.  It is well known \cite{2}
that in the case
of a periodic problem in Euclidean time the derivative of the action with respect to
the energy $E$ is the negative of the oscillation time $\tau(E)$ or period
$P(E)$ at that energy which has
to be identified with $-\hbar/T$.  If $\tau(E)$ is a monotonically decreasing
function with increasing $E$, one has a second order transition; if $\tau(E)$ has a minimum
and then rises again within the domain $0 < E <$ barrier height,
one has a first order transition, i.e. in this case there is 
an energy $E_c$ within this range with $T_c=\hbar/\tau(E_c)$
at which the first order transition takes place.
A criterion for a first order transition can be obtained
by studying the Euclidean time period in the neighbourhood of
the sphaleron configuration at the peak of the
potential barrier, i.e. at the bottom of the well of the inverse potential, as
advocated in ref.\cite{2}.  If the
frequency of oscillation about the sphaleron point is $\omega_s$ and oscillations
different from $\omega_s$ about it are possible, a first order
transition requires $\omega^2 > \omega^2_s$ or $\tau - \tau_s < 0$. Thus
this is a condition one can use as a criterion for a first order transition.
Explicit and solvable examples
of both types of transitions in quantum mechanical models have been given recently
in refs.\cite{3,4,5,6}.  It was shown there that simple models, such as those of the 
$1+1$ dimensional scalar field theories with the familiar double well or periodic
or sine--Gordon potentials, do not lead to first order transitions.      
However, as shown in ref. \cite{3}, this changes if the mass is allowed to become
field dependent, a case which occurs, for instance, in
spin tunneling theories, or
if an externally applied field is introduced as      
in refs.\cite{4,5,6,7}, as again in the context of spin tunneling phenomena. The
introduction of such extra dependences implies, of course, the introduction
of appropriate parameters.  Thus in ref.\cite{3} it was shown, for example,
that in the case of a periodic potential and an effective mass with a field
dependence multiplied by a parameter $\lambda$, the period $P(E)$ is
monotonically decreasing for $\lambda < \frac{1}{2}$, but for $\lambda > \frac{1}{2}$
it has a minimum and then rises again, and so leads to a first order
transition.  It is therefore of interest to find criteria for the 
determination of domains of such parameter values in which they lead to one type of
phase transition or the other.  A significant step forward in this
direction on the basis of higher dimensional arguments, was 
achieved in ref.\cite{2}, where an expansion about the time independent
saddle point configuration is considered.  However, many considerations
of basic interest are best illustrated by quantum mechanical models,
which serve as useful prototypes for more complicated theories, and it is natural
to enquire about criteria for the occurrence of first order transitions
in these, and hence to verify the conditions
obtained heuristically as in the models referred to above \cite{3,4,5}. Thus in the
following we consider a scalar theory with a field dependent mass and suitable 
nonlinear potentials.  Expanding the static field (quantum mechanically a
coordinate) about the constant sphaleron configuration
(at the top of the barrier) and proceeding to second order
perturbation theory, we obtain the deviation of
the frequency from its sphaleron value (lowest order perturbation theory)
and hence the criterion for a first order transition.  We then consider various models
considered earlier \cite{3,4,5,6,7} and demonstrate the agreement of the
prediction of the criterion with the results obtained there heuristically.
We add that theories with effective field dependent masses are not unusual; typical
examples are, for instance, Skyrme theories \cite{8}. The instability of the sphaleron
configuration at the top of the barrier in various familiar models has been
shown, for instance, in \cite{9}.

\section{The condition for a first order transition}
We consider a Euclidean action integral of the form
\begin{equation}
S=\int d\tau\left[\frac{1}{2}M(q){\dot q}^2+V(q)\right]
\label{1}
\end{equation}
with Euler--Lagrange equation
\begin{equation}
M(q)\ddot{q}+\frac{1}{2}\frac{\partial M(q)}{\partial q}{\dot q}^2 = \frac{\partial V(q)}{\partial q}
\label{2}
\end{equation}
and $q(\tau +P)=q(\tau)$, where $P=\frac{1}{T}$ is
the period and $T$ the temperature. Since the solutions near the sphaleron solution at the
top of the barrier shown in Fig.1 have information on the order of the ``phase''
transition between quantum and thermal activity regimes, we confine ourselves
to this region in solution space.
We define (as for a maximum of $V(q)$ at $q_s$)
\begin{equation}
-\omega_s^2 \equiv \frac{V^{\prime\prime}(q_s)}{M(q_s)}
\label{3}
\end{equation}
and set in eq.(\ref{2})
\begin{equation}
q = q_s + \eta(\tau)
\label{4}
\end{equation}
where $q_s$ is the sphaleron position with
$V^{\prime}(q_s)=0$. 
Expanding $M(q_s+\eta (\tau))$ and $V(q_s+\eta (\tau))$ and their derivatives in powers of
$\eta (\tau)$, we obtain the fluctuation equation
\begin{eqnarray}
&&\left[M(q_s)+M^{\prime}(q_s)\eta (\tau)+\frac{1}{2}M^{\prime\prime}(q_s)\eta^2(\tau)\cdot\cdot\cdot
\right]{\ddot \eta}(\tau)\nonumber\\
&&+\frac{1}{2}\left[M^{\prime}(q_s)+M^{\prime\prime}(q_s)\eta (\tau)
+\frac{1}{2}M^{\prime\prime\prime}(q_s)\eta^2(\tau)
+\cdot\cdot\cdot\right]\dot{\eta}^2(\tau)\nonumber\\
&&=V^{\prime\prime}(q_s)\eta (\tau) +
\frac{1}{2}V^{\prime\prime\prime}(q_s)\eta^2(\tau) + \frac{1}{6}V^{\prime\prime\prime\prime}
(q_s)\eta^3(\tau)+\cdot\cdot\cdot
\label{5}
\end{eqnarray}
We are interested in small fluctuations $\eta(\tau)$.
The first order equation
\begin{equation}
M(q_s)\ddot{\eta}(\tau) - V^{\prime\prime}(q_s)\eta (\tau) = 0
\label{6}
\end{equation}
has the even solution
\begin{equation}
\eta(\tau) = a\cos(\omega_0\tau),\;\;\; \omega_0^2=\omega_s^2
\label{7}
\end{equation}
This is the solution we choose for the following reason, which we explain for simplicity
for the case $M(q) = 1$.  The classical equation
$$
{\ddot q} = \frac{\partial V}{\partial q}
$$
is invariant under time translations $\tau\rightarrow\tau +$ const. The operator
of the fluctuation equation or second variational derivative of the action $S$ at
the classical configuration $q_c$ must therefore 
possess a zero eigenvalue with eigenfunction $\frac{dq_c}{d\tau}$.  This operator
is also obtained by differentiating the classical equation and implies the
fluctuation equation
$$
\left(\frac{d^2}{d\tau^2} - V^{\prime\prime}(q_c)\right)\psi_i = \omega_i^2\psi_i
$$
If $q_c$ is the constant $q_s$, and if this is a saddle point, we must have (at and around $q_s$)
$\omega_1^2 = 0, \psi_1$ odd (the first excited state),
and $\omega_0^2 < 0$, and $\psi_0$ even (the ground state).  Thus in this case
$\psi_0$ is proportional to $\cos(\omega_0\tau)$. This explains our ansatz (\ref{7})
which is the component of a general fluctuation in the direction of
the negative eigenmode of the fluctuation equation.
The fluctuation equation for the case $M(q) \neq $const. has been discussed
in \cite{10}.

For the solution in the next order of perturbation theory we set
\begin{equation}
\eta(\tau) = a\cos(\omega\tau) + a^2\eta_1(\tau), \;\;\;\;\omega^2=\omega_0^2 + a\triangle_1\omega^2
\label{8}
\end{equation}
We insert this again into eq.(\ref{5}) and retain only terms up to and including those of $O(a^2)$.
Also we reexpress powers of $\cos(\omega\tau)$ in terms of $\cos(n\omega\tau)$
where $n$ is an integer
and use eq.(\ref{6}). Then
\begin{eqnarray}
&&a^2\left[M(q_s)\frac{d^2}{d\tau^2} - V^{\prime\prime}(q_s)\right]\eta_1(\tau)\nonumber\\
&=&a^2\triangle_1\omega^2M(q_s) \cos(\omega_0\tau)\nonumber\\
&+&a^2\left[\left(\frac{\omega^2M^{\prime}(q_s)+V^{\prime\prime\prime}(q_s)}
{4}\right)+\left(\frac{3\omega^2M^{\prime}(q_s)}{4}+\frac{V^{\prime\prime\prime}(q_s)}{4}
\right)\cos(2\omega_0\tau)\right]
\label{9}
\end{eqnarray}
We now expand the fluctuation $\eta_1(\tau)$ in terms of lowest order functions, i.e.
we set 
$$
\eta_1(\tau)=\sum_{n=0,1,2,\cdot\cdot\cdot}c_n\cos(n\omega_0\tau)
$$
and use the orthonormality of the latter, i.e.
$$
\int^{\frac{P}{2}}_{-\frac{P}{2}}\cos(m\omega_0\tau)\cos(n\omega_0\tau) d\tau
=\frac{P}{2}\delta_{mn}, \;\;\; P=\frac{2\pi}{\omega_0}
$$
The remaining steps are standard in perturbation theory.  Thus, multiplying eq.(\ref{9}) by
$\cos(\omega_0\tau)$ and integrating one obtains immediately that
$$
\triangle_1\omega^2 = 0
$$
and the fluctuation $\eta_1(\tau)$ is obtained to be
\begin{equation}
\eta_1(\tau) = g_1 + g_2\cos(2\omega_0\tau)
\label{10}
\end{equation}
with
\begin{equation}
g_1(\omega)= -\frac{\omega^2M^{\prime}(q_s)+V^{\prime\prime\prime}(q_s)}
{4V^{\prime\prime}(q_s)},\;\;\;
g_2(\omega)= -\frac{3M^{\prime}(q_s)\omega^2+V^{\prime\prime\prime}(q_s)}
{4 \left[4M(q_s)\omega_0^2+V^{\prime\prime}(q_s)\right]}
\label{11}
\end{equation}
where in these first order expressions $\omega^2 = \omega_0^2$. 
Thus, in order to obtain a nonvanishing deviation from the sphaleron value, we
have to proceed to the next order of perturbation theory. Hence we set
\begin{equation}
\eta(\tau)=a\cos(\omega\tau) + a^2\eta_1(\tau) + a^3\eta_2(\tau), \;\;
\omega^2 = \omega_0^2 + a\triangle_1\omega^2 + a^2\triangle_2\omega^2
\label{12}
\end{equation}
and insert this into eq.(\ref{5}) and expand up to and including terms of $O(a^3)$.  
The procedure is the same as before but the calculations are now much more involved,
so that we can only cite a main intermediate result.  The equation corresponding to
eq.(\ref{9}) above and multiplied by $\cos(\omega_0\tau)$ and integrated over 
yields the equation
\begin{eqnarray}
&&-aM(q_s)a^2\triangle_2\omega^2\frac{P}{2}
=a^3\Bigg[V^{\prime\prime\prime}(q_s)\left(\frac{g_1}{2}+\frac{g_2}{4}\right)P
+\frac{1}{6}V^{\prime\prime\prime\prime}(q_s)\frac{3}{8}P\nonumber\\
&&+4M^{\prime}(q_s)\omega^2g_2\frac{P}{4}
+M^{\prime}(q_s)\omega^2
\left(\frac{g_1}{2}+\frac{g_2}{4}\right)P
+\frac{1}{2}M^{\prime\prime}(q_s)\omega^2\frac{3}{8}P\nonumber\\
&&-2M^{\prime}(q_s)
\omega^2g_2\frac{P}{4}
-\frac{1}{2}M^{\prime\prime}(q_s)\omega^2\frac{P}{8}\Bigg]
\label{13}
\end{eqnarray}
From this the condition of a first order phase transition, i.e. $\triangle_2\omega^2 > 0$,
becomes
\begin{eqnarray}
&&\Bigg[V^{\prime\prime\prime}(q_s)\left(g_1+\frac{g_2}{2}\right)
+\frac{1}{8}V^{\prime\prime\prime\prime}(q_s)
\nonumber\\
&&+M^{\prime}(q_s)\omega^2g_2 + M^{\prime}(q_s)\omega^2\left(g_1+\frac{g_2}{2}\right)
\nonumber\\
&&+\frac{1}{4}M^{\prime\prime}(q_s)\omega^2\Bigg]_{\omega_0} < 0
\label{14}
\end{eqnarray}
The first two terms of the expression on the left can be seen to be
contained in the appropriate formula of ref.\cite{2} (there eq.(31)) where $M =$ const.

\section{Application of the criterion to various models}
In the following we focus our interest on the criterion for a first order transition in various
model theories and not on the physical contexts of these theories.  We therefore do not
elaborate on the latter which are discussed in the appropriate sources cited. 

\begin{enumerate}

\item

{\it A spin tunneling model with variable mass but without applied magnetic field}

\noindent
In ref.\cite{3} a large--spin tunneling model was considered in which
\begin{equation}
M(\phi) = \frac{1}{2K_1\left(1-\lambda\sin^2\phi\right)},\;\;\;\;
V\left[\phi\right]=K_2s(s+1)\sin^2\phi
\label{15}
\end{equation}
Thus, identifying $\phi$ with our previous coordinate $q$, we have
$$
\phi_s = \frac{\pi}{2},\;\; V\left[\phi_s\right]=K_2s(s+1),\;\; V^{\prime}\left[\phi_s\right]=0,\;\;
$$
$$
V^{\prime\prime}\left[\phi_s\right]=-2K_2s(s+1),\;\; V^{\prime\prime\prime}\left[\phi_s\right]=0,\;\;
V ^{\prime\prime\prime\prime}\left[\phi_s\right]=8K_2s(s+1)
$$
and
$$
M\left[\phi_s\right]=\frac{1}{2K_1(1-\lambda)},\;\;M^{\prime}\left[\phi_s\right]=0,\;\;
M^{\prime\prime}\left[\phi_s\right]=-\frac{\lambda}{K_1(1-\lambda)^2}
$$
From eq.(\ref{3}) we then obtain
\begin{equation}
\omega_0^2=4K_1K_2s(s+1)(1-\lambda)
\label{16}
\end{equation}
It follows that $g_1=0, g_2=0$, and eq.(\ref{14}) becomes
$$K_2s(s+1) + \left(-\frac{\lambda}{K_1}\right)\frac{1}{(1-\lambda)^2}
K_1K_2s(s+1)(1-\lambda) < 0,
$$
i.e.
\begin{equation}
\lambda > \frac{1}{2}
\label{17}
\end{equation}
This is precisely the condition found heuristically in ref.\cite{3} for the existence of a
first order transition in the model. The result implies, of course, the necessity of
a significant field dependence of the effective mass to generate the first order
transition. In the next model this effect is produced by an applied magnetic
field.

\item

{\it A spin tunneling model with constant mass and applied magnetic field}
  
In ref.\cite{6} a large--spin tunneling case was considered in which (in our present notation)
\begin{equation}
M = const., \;\; V\left[\phi\right]=(s+\frac{1}{2})^2D\left(h_x^2\sinh^2\phi-2h_x\cosh\phi\right)
\label{18}
\end{equation}
Here $h_x=\frac{H_x}{(2s+1)D}$, where $H_x$ is the magnetic field applied in the
direction of coordinte $x$ which in our notation is also $\phi \equiv x$.
In this case one finds
$$
 \phi_s =0, \;\;V\left[\phi_s\right]=-2(s+\frac{1}{2})^2Dh_x, \;\; V^{\prime}\left[\phi_s\right] = 0,
$$
$$
V^{\prime\prime}\left[\phi_s\right]= -2(s+\frac{1}{2})^2h_x(1-h_x), \;\;
V^{\prime\prime\prime}\left[\phi_s\right] = 0,
$$
$$
V^{\prime\prime\prime\prime}\left[\phi_s\right]=2(s+\frac{1}{2})^2Dh_x(4h_x-1)
$$
and
$$
M = M, M^{\prime}=M^{\prime\prime}=0
$$
Then from eq. (\ref{3})
\begin{equation}
\omega_0^2=\frac{2}{M}(s+\frac{1}{2})^2h_x(1-h_x) 
\label{19}
\end{equation}
and $g_1 =0, g_2 = 0$ and the inequality (\ref{14}) becomes
$$
\frac{1}{8}\cdot 2(s+\frac{1}{2})^2Dh_x(4h_x-1) < 0
$$
and hence
\begin{equation}
h_x < \frac{1}{4}
\label{20}
\end{equation} 
as found in ref.\cite{6}, there expressed as $H_x < sD/2$.

\item

{\it A spin tunneling model with variable mass and applied transverse magnetic field}

In ref.\cite{4} another and more complicated spin tunneling model was investigated
in which the magnetic field is applied in a direction transverse to
that of the easy axis and the
potential has the shape of a periodically recurring asymmetric twin barrier
(i.e. a large one followed by a small one) with
\begin{eqnarray}
M(\phi) &=& \frac{1}{2K\left(1-\lambda\sin^2\phi + \alpha\lambda\sin\phi\right)},\nonumber\\
V\left[\phi\right] &=& K\lambda s^2\left(\sin\phi - \alpha\right)^2
\label{21}
\end{eqnarray}
In this case we have to distinguish between transitions in the large barrier and
those in the small barrier. We therefore consider these separately.

{\it (i) Large barrier transitions}

In this case the sphaleron angle is (superscript $L$ referring to the large barrier)
$$
\phi^L_s =\frac{3}{2}\pi 
$$
and there
$$
V\left[\phi^L_s\right]=K\lambda s^2(1+\alpha)^2,\;\;V^{\prime}\left[\phi^L_s\right] = 0,
$$
$$V^{\prime\prime}\left[\phi^L_s\right]=-2K\lambda s^2(1+\alpha),\;\;
V^{\prime\prime\prime}\left[\phi^L_s\right] = 0,\;\;
V^{\prime\prime\prime\prime}\left[\phi^L_s\right] = 2K\lambda s^2(4+\alpha)
$$
and
$$
M(\phi^L_s) = \frac{1}{2K\left[1-\lambda\left(1+\alpha\right)\right]},\;\;
M^{\prime}(\phi^L_s) = 0,\;\;
$$
$$
M^{\prime\prime}(\phi^L_s)=-\frac{\lambda}{2K}\frac{(2+\alpha)}
{\left[1-\lambda\left(1+\alpha\right)\right]^2}
$$
In this case we obtain with eq. (\ref{3})
\begin{equation}
{\omega^L_0}^2 = 4K^2\lambda s^2(1+\alpha)\left[1-\lambda\left(1+\alpha\right)\right]
\label{22}
\end{equation}
Again $g_1=0, g_2=0$ and the condition for a first order phase transition
is found to be
\begin{equation}
\lambda > \frac{\alpha + 4}{\left(\alpha + 1\right)\left(3\alpha + 8\right)}
\label{23}
\end{equation}
The parameter domains for first order and second order transitions through the large barrier
are shown in Fig. 2.

{\it (ii) Small barrier transitions}

In the case of transitions through the small barrier we have
$ \phi^S_s = \frac{\pi}{2}$ and
$$
V\left[\phi^S_s\right] = K\lambda s^2 \left(1-\alpha\right)^2,\;\;
V^{\prime}\left[\phi^S_s\right] = 0, \;\; V^{\prime\prime}\left[\phi^S_s\right]
=-2K\lambda s^2\left(1-\alpha\right),
$$
$$
V^{\prime\prime\prime}\left[\phi^S_s\right] = 0, 
V^{\prime\prime\prime\prime}\left[\phi^S_s\right]= 2K\lambda s^2\left(4-\alpha\right)
$$
and 
$$
M(\phi^S_s) = \frac{1}{2K\left[1-\lambda\left(1-\alpha\right)\right]}, \;\;
M^{\prime}(\phi^S_s) = 0,
$$
$$
M^{\prime\prime}(\phi^S_s)=-\frac{\lambda}{2K}\frac{\left(2-\alpha\right)}
{\left[1-\lambda\left(1-\alpha\right)\right]^2}
$$
From eq. (\ref{3}) we find
\begin{equation}
{\omega^S_0}^2 = 4K^2\lambda s^2\left(1-\alpha\right)\left[1-\lambda\left(1-\alpha\right)\right]
\label{24}
\end{equation}
Again $g_1 = 0, g_2 = 0$ and the condition for a first order transition through the small
barrier is found to be
\begin{equation}
\lambda > \frac{4-\alpha}{\left(8-3\alpha\right)\left(1-\alpha\right)}
\label{25}
\end{equation}
The parameter domains for first and second order transitions through the small
barrier are shown in Fig. 3. 
The parameter domains for transitions through either barrier are shown in Fig. 4.

\item

{\it A spin tunneling model with variable mass and applied longitudinal field}

In ref.\cite{7} yet another model was investigated in which the applied magnetic field 
assumes a longitudinal, i.e. parallel, direction with respect to the easy axis.  
In this case it is also possible to find explicit classical configurations \cite{11}, but we
do not consider these here.  In this model \cite{7} 
\begin{eqnarray}
M(\phi) &=& \frac{1}{2K_1\left(1-\lambda\sin^2\phi-\frac{\alpha\lambda}{2}\cos\phi\right)},\nonumber\\
V\left[\phi\right] &=& K_2s^2\left(\sin^2\phi+\alpha\cos\phi+\alpha\right)
\label{26}
\end{eqnarray}
In this case the sphaleron value of $\phi$ is $\phi_s=\arccos\frac{\alpha}{2}$ and
$$
V\left[\phi_s\right]=K_2s^2\left(1+\alpha+\frac{\alpha^2}{4}\right),\;\;
V^{\prime}\left[\phi_s\right] = 0
$$
$$
V^{\prime\prime}\left[\phi_s\right]=-K_2s^2\left(2-\frac{\alpha^2}{2}\right),\;\;
V^{\prime\prime\prime}\left[\phi_s\right]=-3K_2s^2\alpha\sqrt{1-\frac{\alpha^2}{4}},
$$
$$
V^{\prime\prime\prime\prime}\left[\phi_s\right]=K_2s^2\left(8-\frac{7}{2}\alpha^2\right)
$$
and 
$$
M(\phi_s)=\frac{1}{2K_1\left(1-\lambda\right)},\;\;M^{\prime}(\phi_s)=
\frac{\lambda}{4K_1}\frac{\alpha\sqrt{1-{\frac{\alpha^2}{4}}}}{\left(1-\lambda\right)^2},
$$
$$
M^{\prime\prime}(\phi_s)=\frac{\lambda}{2K_1}\frac{\left(\frac{3}{4}\alpha^2 - 2\right)
\left(1-\lambda\right)+\frac{\alpha^2\lambda}{2}\left(1-\frac{\alpha^2}{4}\right)}
{\left(1-\lambda\right)^3}
$$
In this case
\begin{equation}
\omega_0^2= 2K_1K_2s^2(1-\lambda)\left(2-\frac{\alpha^2}{2}\right)
\label{27}
\end{equation}
The expressions for $g_1$ and $g_2$ are now found to be
\begin{eqnarray}
g_1&=&\frac{\alpha}{8\sqrt{1-\frac{\alpha^2}{4}}}\left[\frac{\lambda}{(1-\lambda)}
\left(1-\frac{\alpha^2}{4}\right)-3\right],\nonumber\\
g_2&=&-\frac{\alpha}{8\sqrt{1-\frac{\alpha^2}{4}}}\left[\frac{\lambda}{1-\lambda}\left(1-
\frac{\alpha^2}{4}\right)-1\right]
\label{28}
\end{eqnarray}
The condition for a first order phase transition is now more complicated in that it
reduces to the inequality
\begin{equation}
\frac{3}{16}\alpha^2\xi^2 - \xi + \left(1+\frac{\alpha^2}{2}\right) < 0
\label{29}
\end{equation}
where
$$
\xi = \frac{\lambda}{1-\lambda}\left(1-\frac{\alpha^2}{4}\right)
$$
Setting
$$
\xi_{\pm}=\frac{8}{3\alpha^2}\left[1\pm\sqrt{1-\frac{3}{4}\alpha^2\left(1+\frac{\alpha^2}{2}
\right)}\right]
$$
the inequality (\ref{29}) is then $\xi_-<\xi<\xi_+$, which reduces to
\begin{equation}
\frac{\xi_-}{\xi_-+\left(1-\frac{\alpha^2}{4}\right)} < \lambda
< \frac{\xi_+}{\xi_++\left(1-\frac{\alpha^2}{4}\right)}
\label{30}
\end{equation}
In Fig. 5 we plot the boundary separating the parameter domain of first order
transitions from that of second order transitions.

\end{enumerate}

\section{Conclusions}
In the above we have derived a criterion for the parameter domains of first order phase
transitions for a wide class of quantum mechanical tunneling models.  By application to specific
models which received attention recently, particularly in the discussion of the tunneling of
large or macroscopic spins, we demonstrated the agreement of the prediction of
the criterion with
previous heuristically obtained results.  
We envisage that considerations like those presented above and in ref.\cite{2}
can be useful in wider contexts, and also in field theory models, such as those with
Skyrme terms which imply an effective field dependent mass.

\vspace{2cm}

\noindent
{\large \bf Acknowledgements}

D. K. P. acknowledges support of the Deutsche Forschungsgemeinschaft (DFG) and the
Korea Science \& Engineering Foundation (KOSEF), and J. M. S. R. support by a DAAD fellowship.

\newpage
\begin{center}
{\large \bf Figure captions}
\end{center}
\vspace{1.2cm}
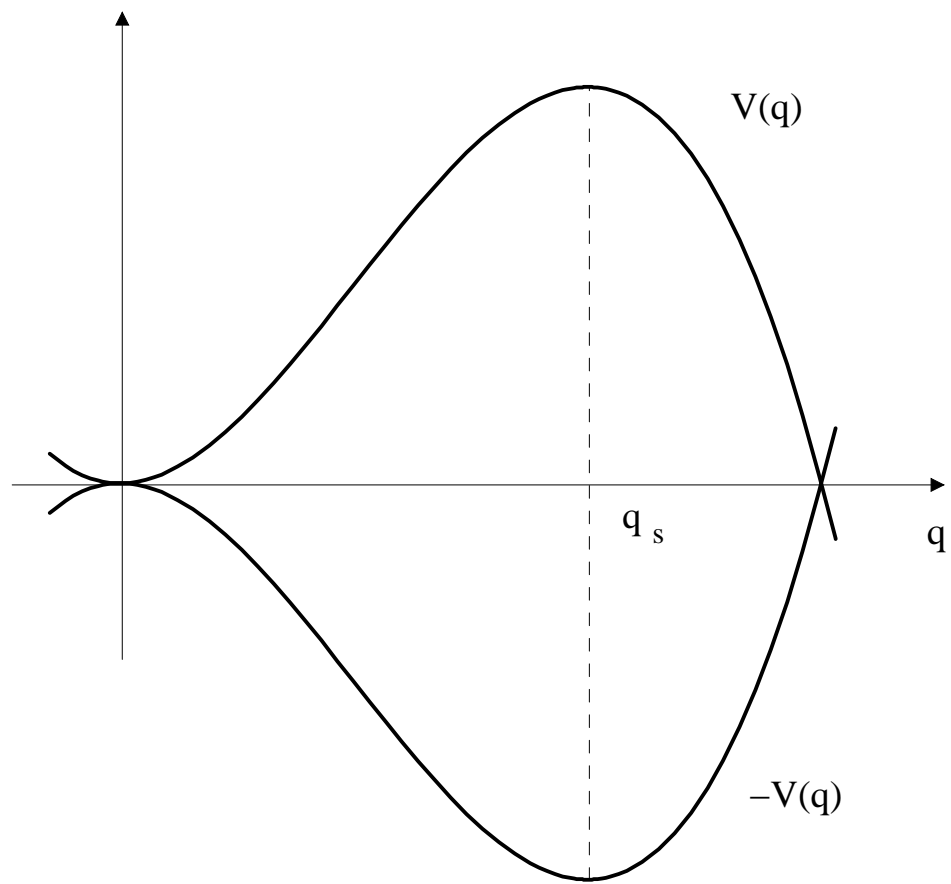
\centerline{\bf Fig.1}

The potential barrier with sphaleron coordinate $q_s$

\vspace{1.2cm}
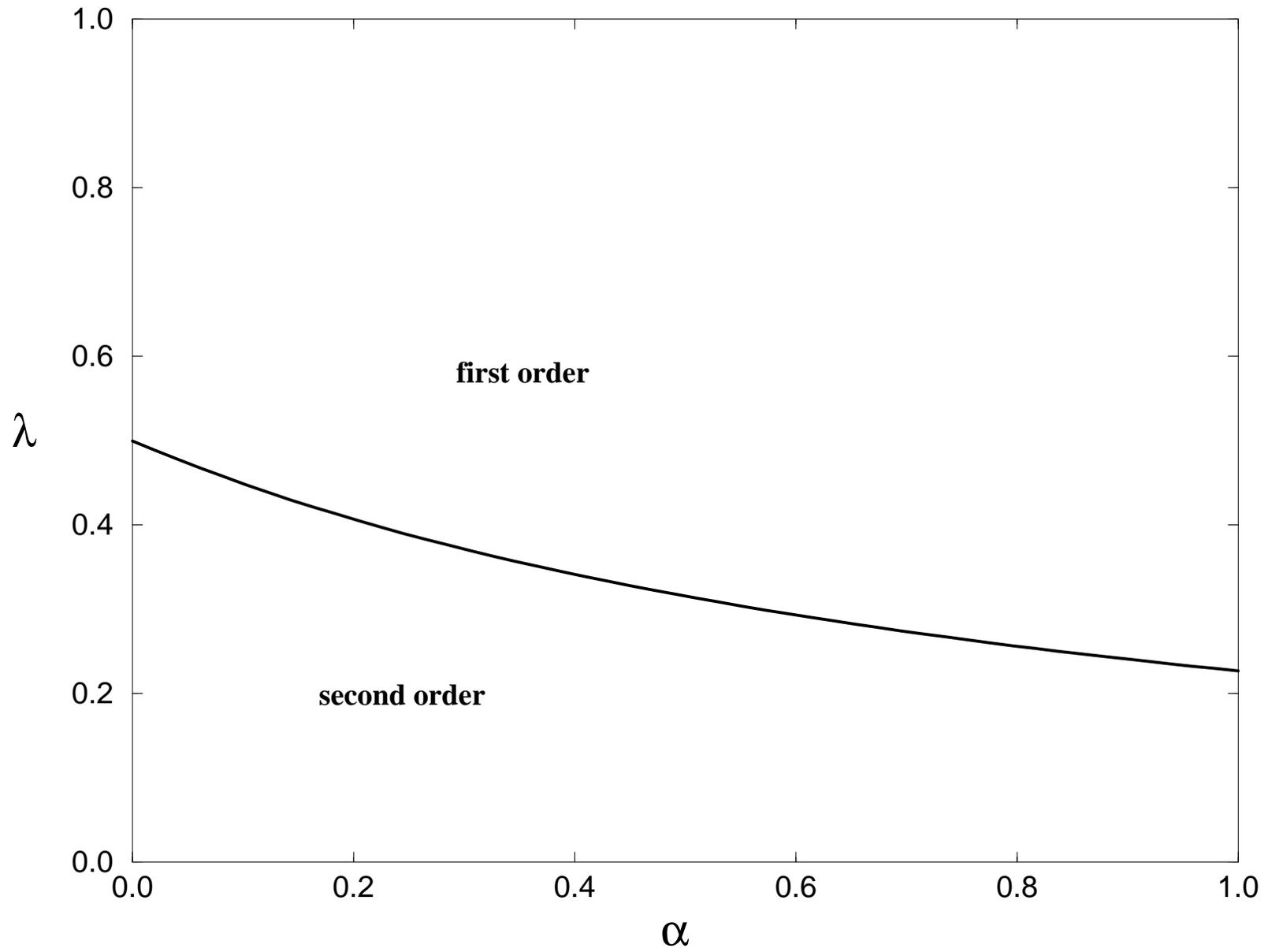
\centerline{\bf Fig. 2}
The parameter domains
for first and second order transitions through the large barrier
in the spin tunneling model with field dependent mass and field transverse
to the easy axis

\vspace{1.2cm}
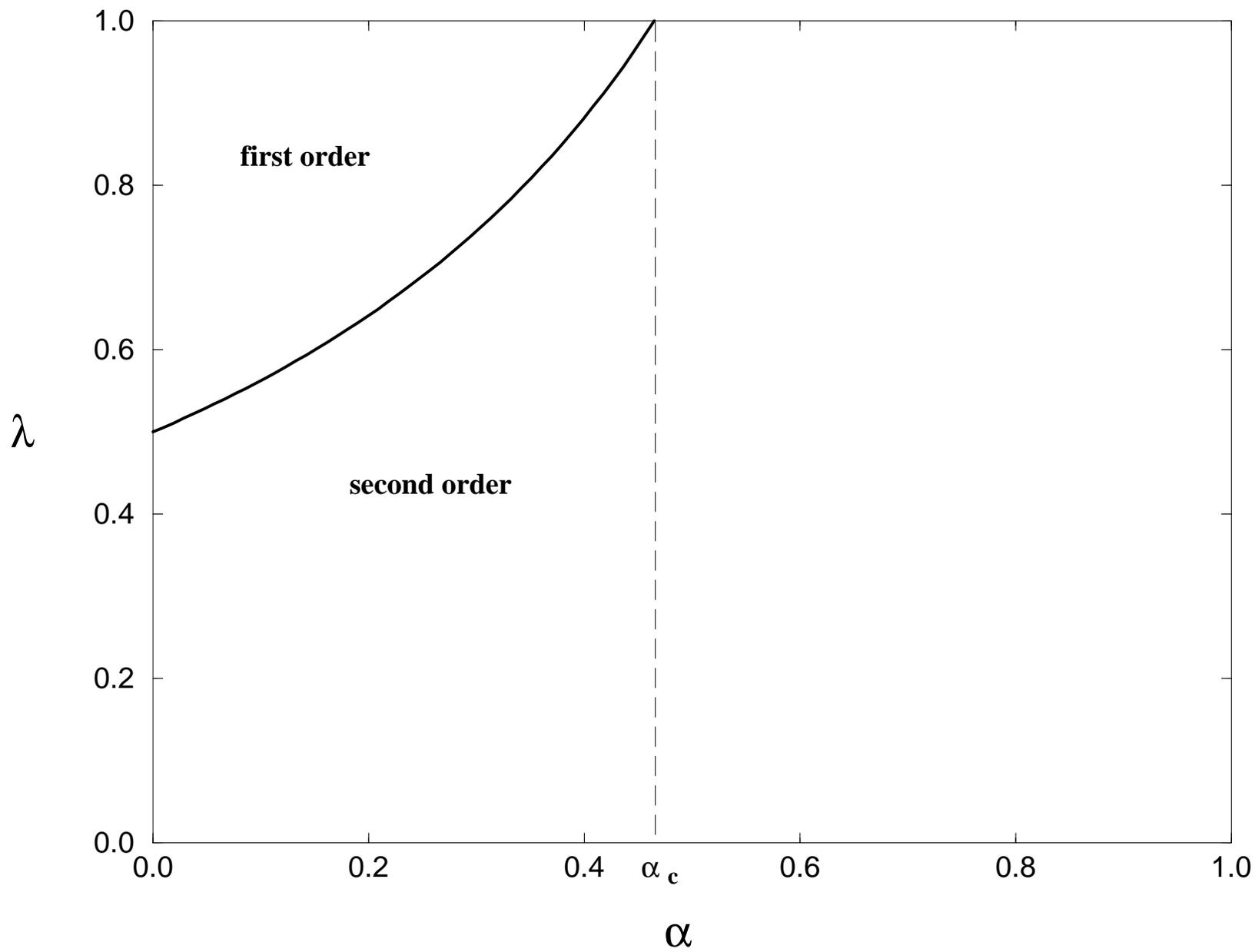
\centerline{\bf Fig. 3}
The parameter domains for first and second order transitions through the small
barrier in the spin tunneling model with field dependent
mass and field transverse to the easy axis

\vspace{1.2cm}
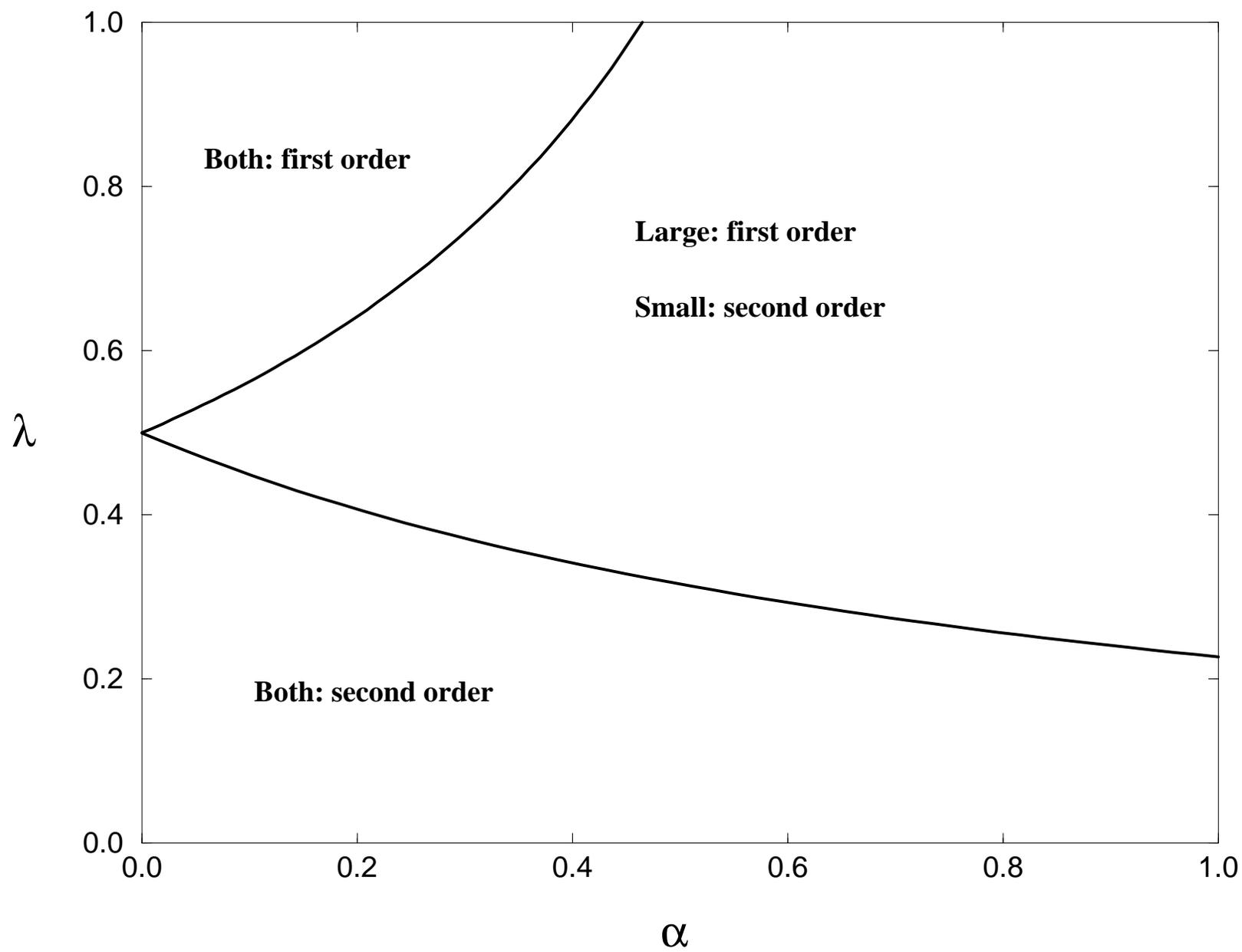
\centerline{\bf Fig. 4}
The parameter domains for first and second order transitions in the model
with field dependent mass and transverse magnetic field

\vspace{1.2cm}
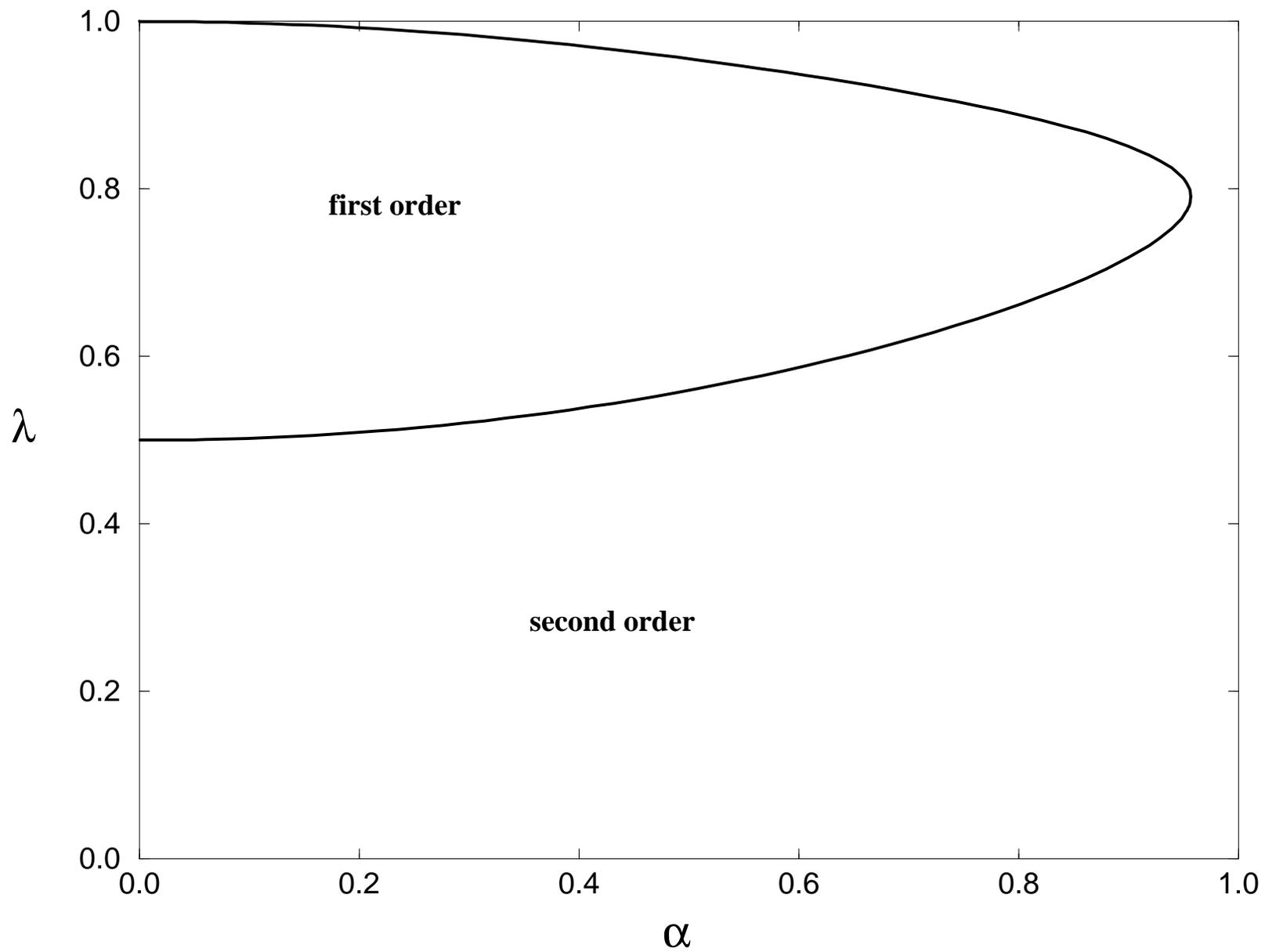
\centerline{\bf Fig. 5}
The parameter domains for first and second order phase transitions in the
model with applied longitudinal field

\end{document}